# Efficient solar collection beyond the étendue limit

Rakan E. Alsaigh,[1,*] Ralf Bauer,[2] and Martin P.J. Lavery [1,*]

[1]School of Engineering, University of Glasgow, Glasgow, UK
[2]Department of Electronic and Electrical Engineering, University of Strathclyde, Glasgow, UK
*Corresponding authors. Emails: r.alsaigh.1@research.gla.ac.uk and martin.lavery@glasgow.ac.uk

**Photovoltaics (PV) are a versatile and compact route to harness solar power. One critical challenge with current PV is persevering the optimal panel orientation angle with respect to the sun for efficient energy conversion. We experimentally demonstrate a bespoke array of multi-element telecentric optical lenses that allow for a greatly increased open circuit voltage ($V_{oc}$) at solar incidence angle of between ±80 degrees compared to a standard panel. Our prototype lens-let shows a 132% increase in $V_{oc}$ over a full day, at optical incidence angles of ±80 degrees. This many element array provides increased field of view without breaking the conservation of étendue. Our prototype indicates that this lens-let array could potentially be mass produced and be readily installed onto any PV system.**

# Introduction

Solar photovoltaic is a renewable technology with a vast potential to drive a sustainable future [1]. This technology does not only play an important role in supporting the distributed generation of electricity around the world, it also offers a route to power many mobile devices, autonomous drones, spacecrafts, domestic rooftops and off-the-grid communities. Unfortunately, only a fraction of solar energy can be efficiently collected by photovoltaic systems due to the physical constraints of collecting solar energy over a wide acceptance angle, photon conversion and charge generation limitations in the absorbing layer within the semiconductor material [2–4]. Novel optical elements are an interesting and powerful route to address and resolve some of these challenges.

Novel optical components are common in concentrated photovoltaic (CPV) [5–7]. These systems utilize large optical collection systems to focus solar radiation onto a small photovoltaic panel, based on the principle that PV is more expensive than large optical components [8–14]. In these concentrated solar power solutions, one makes a critical trade-off by increasing the power per unit area at the back-focal plane of the optical system through limiting its field of view (FOV), where state of the art concentrating photovoltaic have a typical field of view of ≤1° and therefore require precise sun trackers that can be constrained to a tolerance less than 1° for high concentration enhancement [15–17]. The strategic investments over the last decade in the manufacturing of non-concentrating PV system have drastically changed the solar power industry, where the cost of deploying PV cells is now lower than concentrated power collection and in many areas of the world provides power at a cost-per-Watt that is cheaper than fossil fuels [18].

Traditionally, the angle-dependent efficiency constraints in PV have been tackled through the use of mechanical tracking systems that are expensive, require continual maintenance and the use of sensors to create automatic systems. Static systems offer a clear advantage over mechanical systems in-terms of maintenance, fault tolerance, and installation. Many of the current leading passive optical solutions for this challenge require fundamental changes to industry manufacturing processes, which include light trapping techniques [19], surface coating [20] or surface integration of novel components [8]. However, given the maturity of PV production, substantial changes to manufacturing techniques are not ideal. A laudable goal is the development of external non-concentrating optical components that can be readily added to standard PV panels. Optical elements suitable for mass production could be widely deployed to increase their full day collection efficiency, or for applications when there is rapidly changing or no fixed angular relationship with the sun.

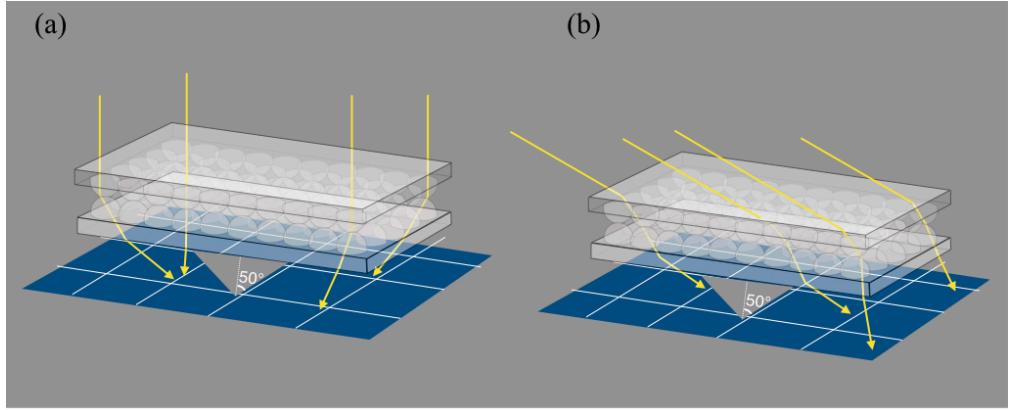

**Fig. 1.** Ultra-wide angle optical element. (a) Schematic illustration of the system arrangement, with the optical element on-top of a PV panel. The optical element consists of two face-to-face layers of 2mm diameter hemispheric lens-let arrays that sit on a 2mm thick base. Simulated incident rays are displayed at 0° angle of incidence with transmitted rays at angles up to 50°. The angle of the transmitted rays is largely dependent on the position at which the incident rays hit the top hemispheric lens. (b) Schematic illustration of the system with simulated rays at incident angle of 60° and transmitted rays at angles up to 50°.

In this paper we outline the use of a novel non-concentrating static lens-let array that can increase the PV open circuit voltage by more than 132% over a full day, at optical incidence angles of ±80 degrees. For solar panels installed at normal angles to the sun our optical element could remove the requirement to reposition solar panels through the year and greatly increase solar collection efficiency in the summer months where there are longer periods of sunlight. These significant gains could potentially be higher for mobile systems that have no direct alignment towards the sun, such as planes, drones, electric vehicles and robots. Our approach uses two-layers of hemispherical lens-lets array that are engineered to redirected light incident at ±80 degrees to be at an output angle within the field of view of commercially available solar panels, as shown in Fig. 1. We have produced a 3D-printed prototype, shown in Fig. 2(a), that can be readily retrofitted to new or already deployed solar panels. Our prototype indicates that the optical element could potentially be mass produced through mature industrial scale manufacturing techniques such as injection moulding.

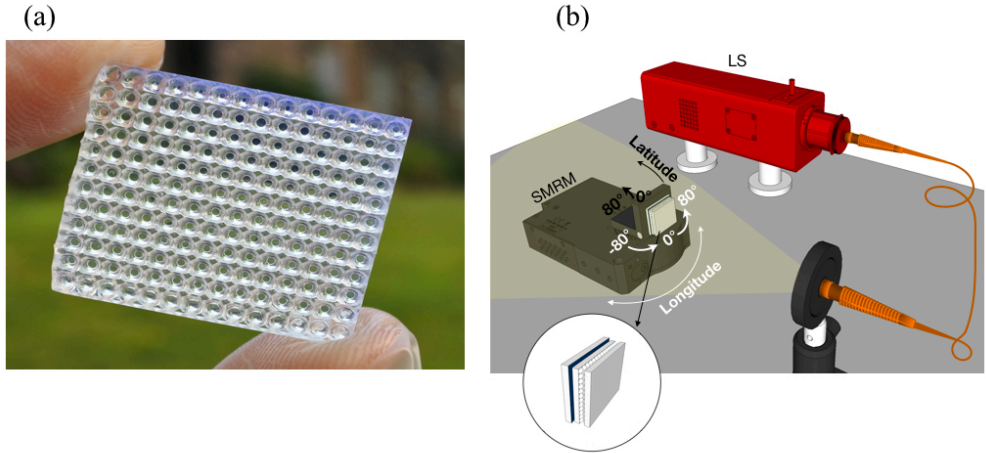

**Fig. 2.** 3D-printed optical element. (a) Image of our 3D-printed optical element. (b) Illustration of the testing set-up, showing a fiber-coupled stabilised broadband light source (LS) illuminating a 15cm-diameter beam at the optical element's plane. The optical element is on-top of an amorphous Si solar panel that is back-attached to a tilt-adjustable rotational mount driven by a stepper motor (SMRM).

## Results

### Optical element design

Our optical element is comprised of two identical 2D arrays of hemispheric lens-lets, where each individual lens-let has a 2 mm diameter and 2 mm aperture. The arrays are positioned and optically bonded with their curved faces touching, as shown in Fig. 1. Each pair of hemispheric lenses within the array forms a telecentric compound lens with a large field of view [21]. The small aperture size (2 mm) and short focal length (2 mm) allow for high incidence angle input rays entering the aperture of the first lens in the pair to be readily collected by the second lens. Forming these miniature hemispheric lenses into a bespoke lens-let array provides the increased field of view from the small form factor telecentric lens, while simultaneously allowing for coverage of large area solar panels without bulky concentrator optics.

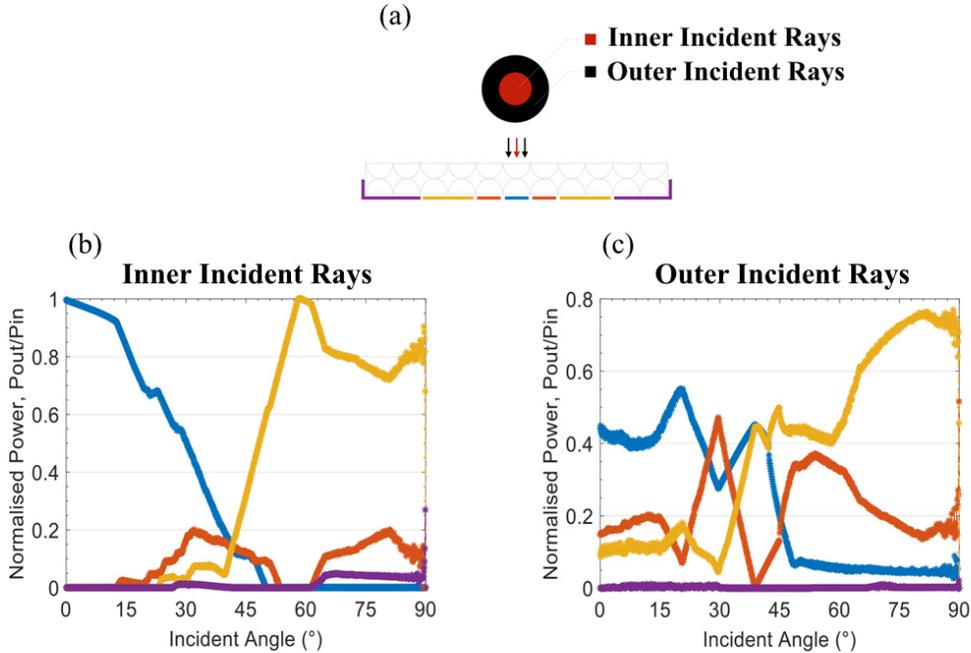

**Fig. 3.** Simulated normalised power. (a) Incident rays into a single hemispheric lens-let are simulated for the inner and outer apertures, which are 33.3% and 66.6% of lens diameter, respectively. The optical power was separately collected at the back plane of the second lens-lets and plotted as a normalised value of the incident power (Pout/Pin). (b) Normalised transmitted power at different back plane positions for rays incident into the centre of the lens-let aperture as a function of 0°-90° longitudinal incident angle. (c) Normalised transmitted power for rays incident into the outer sides of the aperture.

Our two-layer construction of the lens-let array allows for rays that are not initially collected by the second lens in any given pair to be collected by a neighbouring lens. Using the ray tracing model, we calculated the percentage of incident power that couples into the adjacent lens pairs, as shown in Fig. 3. It can be seen that the angle and lateral position of an incoming ray plays a role in the percentage of power coupled into neighbouring elements. This model predicts that the majority of optical power is collected by the solar panel within three lens pairs either side of the lens that any given ray is initial incident on, regardless of their position or angle of incidence.

An ideal optical system for efficient solar collection would redirect input rays from any arbitrary input angles to a single propagation direction. However, conservation of étendue prevents this from happening [22]. A technical trade-off is a system designed to more efficiently collect from the widest possible range, and redirect these rays to be within a smaller range of propagation directions, Fig. 4. A ray traced model of our system was developed and the angular direction with respect to the optical surface normal was calculated for between 0 to 89.5 degrees. Using this data, the percentage of rays found to be below and above 50 degrees incidence at the solar panel surface were calculated, shown in Fig. 4. It can be seen from this simulated data that our current optical design redirects a large percentage of rays with incident angles above 50 degrees to input incidence angles of less than 50 degrees when hitting the solar panels surface. Although 20% of rays at small angles of incidence are directed to be out with the highest efficiency collection angle, we don't observe any noticeable loss of efficiency.

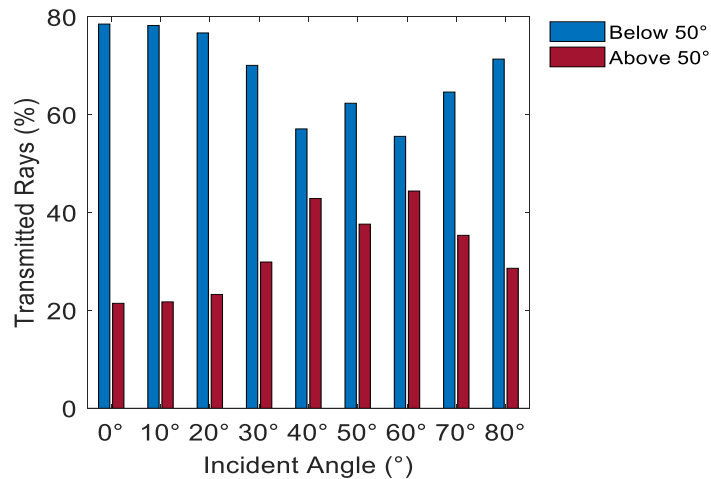

**Fig. 4.** Simulated angular deflection. Percentage of transmitted rays that are below and above 50° are displayed summed over 10 degrees opening angles in steps of 10 degrees between 0 and 80 degrees. The percentage shown is only calculated for the whole of each set of angles, while the number of total simulated rays below and above 50° are 30175 and 14336, respectively.

**Prototype fabrication**

Due to the novel nature of the optical lens-let array, we manufactured our prototype lens through the use of micron-resolution high-precision stereolithographic 3D-printing. Recent advances in 3D-printer technology have enabled the formation of clear photo-polymerised samples, which have been demonstrated in various applications [23–26]. The printing of these materials combined with specific post-processing procedures has improved the surfaces roughness to the nanometer scale [23, 27]. We produced optical quality lens-let arrays using a widely available 3D-printer combined with several stages of post-processing, see Supplementary Fig.S1. Each of our hemispheric lens-let arrays were printed separately and subsequently bonded with medium-viscosity UV-curable index-matched fluid, providing a transparent boundary layer joining the curved faces, see Fig. 1. The optical element was then directly fitted to a 29.05 x 24.3mm standard commercial amorphous silicon solar panel using index-matching fluid to minimise intermediate reflection losses of light at the boundary between our lens-let array and the solar panel. Further, we tested the heat resilience of these plastics in direct solar illumination at ambient temperatures of over 50 degrees Celsius and observed no noticeable damage to the 3D-printed materials.

The transmission loss associated with our 3D-prints was determined by using a broadband white light source. Using a spectrometer, the overall optical loss for our lens-let array was measured to be 16.4%. To determine contribution for absorptive losses versus back reflective losses, we measured the loss from flat surfaced 3D-printed boxes of varying thicknesses. We calculated the plastic absorptive loss to be approximately 1.3% per mm, which likely occurs due to the internal plastic layering, and reflective losses per optical surface of approximately 4.3%. We expect one could mitigate these losses through the use of anti-reflection coated optical surfaces and use of a higher precision 3D-printer or injection moulded plastics. Our measured loss show a slight improvement compared to previous research that employs similar approaches to 3D-printing optics [24, 28].

**Experimental performance**

To experimentally investigate the feasibility of our prototype design, we measured the open circuit voltage ($V_{oc}$) across the solar panel over a wide range of illumination angles using a fiber-coupled stabilised white light source as solar model. Utilising a stepper motor-driven rotation mount, a range of solar illumination angles were tested over the range -80 degrees to 80 degrees in the longitudinal axis, where a further mechanical rotation mount was used to allow for tilt of the solar panel for latitudinal alignment. The solar panel was placed 17.5cm away from the light source to yield uniform plane wave illumination over the full collection aperture and mimic solar radiation, shown in Fig. 2(b)**.**

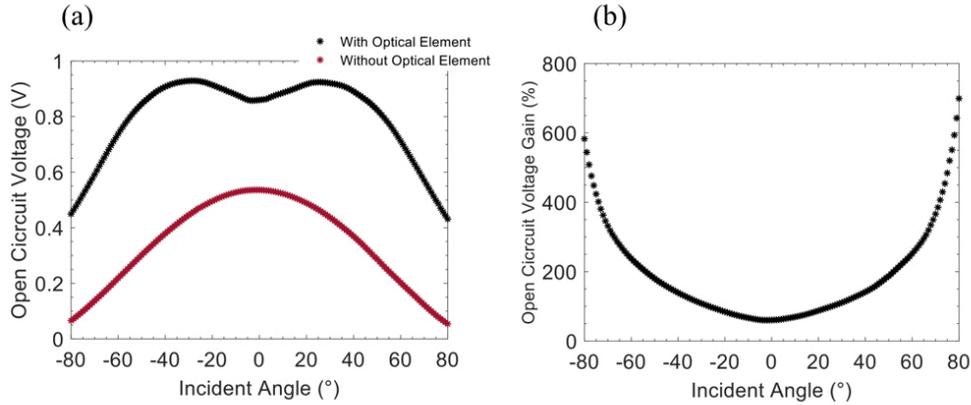

**Fig. 5.** 80° Latitude tilted experimental testing. (a) Open circuit voltage measurement across a bare amorphous Si solar panel (red), with the addition of our 3D-printed optical element on-top of the same solar panel (black) over ±80° longitudinal incident angles, both under the same 80° latitude tilt illumination of a fiber-coupled stabilised broadband light source. (b) Percentage gain in open circuit voltage shown in a for the addition of our optical element over the bare panel.

Our white light source (Thorlabs SLS201L) is designed to provide a black body radiation at a temperature of 2796 K. We fully characterised the performance of the solar panel for our light source over the full range of possible angles of incidence, using a bare solar panel, where $V_{oc}$ measurements were taken in 1 degree of longitudinal steps at a fixed latitude tilt of 80 degrees with a measurement time of 20 seconds at each step, Fig. 5(a). The voltage fluctuations were found to be small with a cumulative standard deviation between the samples of 26.51mV. Subsequently, a measurement over the identical longitudinal sweep is taken using the same panel with our lens-let array bonded to the surface, Fig. 5(a). The percentage gain in $V_{oc}$ for the addition of our lens-let array over the bare solar panel is shown in Fig. 5(b).

The daily changes in sun position are only one of the issues that limit the efficiency of solar technologies. Over the course of the year the suns latitude in the sky changes, leading to a secondary angle dependent efficiency change. Even more importantly, any mobile platforms that intend to use solar power (planes, drones, etc.) will never have a guaranteed solar alignment. To test the applicability of our system for such applications we altered the effective latitude by varying the adjustable mount angle with respect to the incoming light. For a selection of five fixed latitude angles, a further scan over the longitudinal range discussed earlier was taken, where the $V_{oc}$ was measured for the bare and element integrated solar panel, and the gain over the bare panel is calculated, Fig. 6. We measure an increase in full day $V_{oc}$ of 3.98±0.01%, 3.24±0.01%,

9.47±0.01%, 19.71±0.02%, 45.12±0.02% and 132.39±0.05% at latitude angles of 0, 20, 40, 60, 70 and 80 degrees, respectively. We expect the performance of a PMMA injection moulded lens-let arrays with low cost anti-reflective coating will increase these gain further to 5.96%, 5.58%, 12.96%, 33.30%, 64.94% and 169.73% at latitude angles of 0, 20, 40, 60, 70 and 80 degrees, respectively.

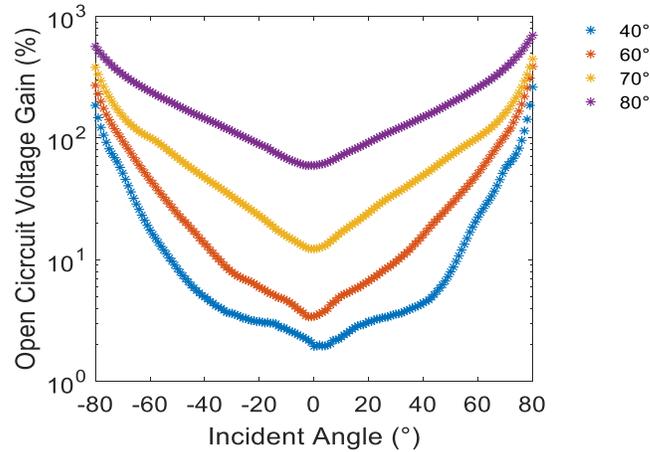

**Fig. 6.** System performance at various tilt angles. Percentage gain in the measured open circuit voltage across the same solar panel used in Fig. 5, at panel latitude tilt angles of 40°, 60°, 70° and 80, as a function of ±80° incident longitudinal angles.

## Discussion

We have developed a novel lens-let arrays that can be used to increase solar collection angles beyond those of traditional étendue conserved optical systems. Given the hemispherical nature of the lens-let array design, mass production is feasible through industrial processes such as injection moulding. Lens-let arrays of this type could be widely used to replace mechanical tracking and allow panels to be more readily fitted to ground or aerial vehicles that have no fixed angle relationship with the sun, such as planes and drones. Large scale deployments of solar panels such as those proposed by Saudi Arabia, a leading oil producing nation, to install very large surface area solar farms based on PV [29] and recent changes in California state regulation [30] that require solar panels to be fitted to every newly build house could be greatly enhanced through the addition of passive optical systems. Further, already deployed solar panels could be readily retrofitted with non-concentrating optical lens-let top layers and could substantially increase their power collection efficiency.

## Materials and Methods

### Ray tracing simulation

Comsol Multiphysics was used to model the optical power transmitted by our optical design. A pair of hemispheric lenses, with the same refractive index of the clear 3D-print material (n=1.5403), was built in Comsol along with two sets of five adjacent double lenses on both sides to form an array of eleven pairs of lens-lets, see Fig. 3(a). The aperture of the central lens pair was segmented into three equal apertures, one inner (central) aperture and two outer (side) apertures, to guide rays through different incident positions and trace the power and position of the transmitted rays using two sets of simulations. A plane wave illumination at λ=660nm was released into the central lens and the transmitted rays were separately collected along the back of the second central lens and the ten adjacent lenses as shown in Fig. 3(a). The accumulated power of these transmitted rays was computed as a function of incident power over incident angles of 0°-90° in a step of 0.1° for both inner and outer apertures as shown in Fig. 3(b) and Fig. 3(c), respectively.

Comsol Multiphysics was also used to create a 2D ray tracing simulation of the whole double lens-let array centred around a cross-section through the largest diameter of the lens-lets. The 2mm diameter hemispherical lens-lets were modelled using the CAD design for 3D-printing, with a 120μm thick index matching layer around the connection points of the two lens-let arrays. The plane wave illumination with 151 rays was traced through the system with 5° orientation steps of the optical element. The ray power at the input plane of the solar panel was accumulated over the full time steps for each orientation, combined with the incident angle of each ray being guided to the solar panel interface. This confirmed the transformation of incident angles into the panel's efficient conversion range below 50°, specifically from solar incident angles at the higher angular range as shown in Fig. 4.

# Experimental lens-let arrays fabrication

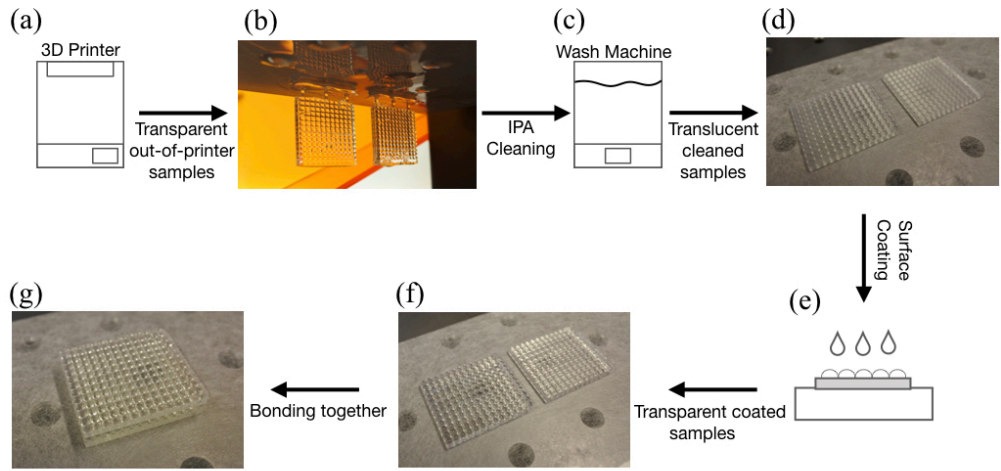

**Fig. S1.** 3D-printed optical element post-processing procedures. (a) Stereolithographic 3D-printer to fabricate the lens-let array. (b) Image of out-of-printer lens-let arrays look transparent due to accumulated resin on the surface. (c) Wash machine filled with Isopropyl alcohol (IPA) to clean the samples. (d) Image of cleaned lens-let arrays as they look translucent due to surface roughness that was previously covered by resin before the cleaning. (e) Surface coating with index-matched material to smoothen the surface and makes it transparent. (f) Post-coating image of transparent lens-let arrays. (g) Lens-let arrays are tip-bonded together to form the optical element.

Supplementary Fig. S1 illustrates the steps taken to 3D-print and post-process the lens-let arrays. A stereolithography (SLA) 3D-printer was used to fabricate the lens-let array from a liquid resin using a $\lambda=405nm$ high power laser in a layer-by-layer photopolymerisation process at layer thickness of 50μm. Due to the nature of SLA printing that forms solid elements from liquid material, the lens-let arrays were cleaned in 90% isopropyl alcohol (IPA) to remove accumulated uncured resin and other residues from the surface. The clean surface was coated with UV-curable index-matching material to smoothen the surface and enhance its transparency. The lens-let arrays were tip-bonded and cured using a $\lambda=405nm$ UV light to form the optical element.

## Loss characterisation

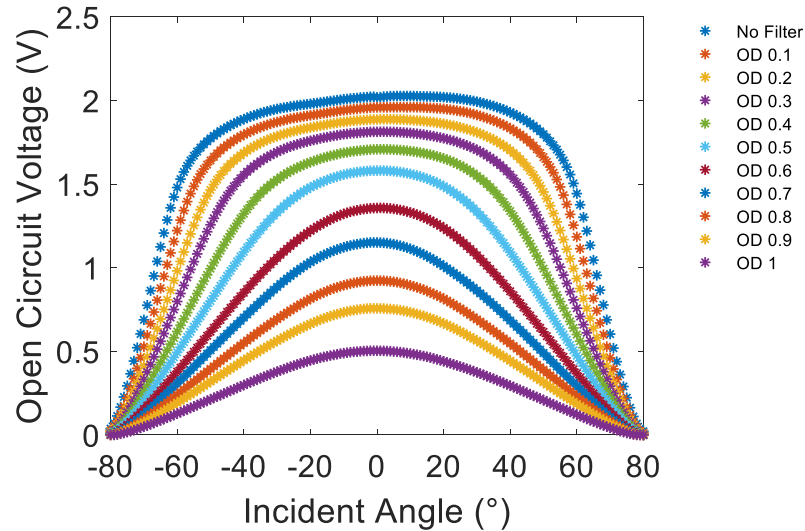

**Fig. S2.** System evaluation with Neutral Density (ND) filters. Open circuit voltage measured as function of incident angle at different optical density from 0.1 to 1 with a step of 0.1.

Neutral Density (ND) filters were used to evaluate the performance of the amorphous silicon solar panel under different optical intensities as well as its response to these changes over a range of longitudinal angles. The optical density (OD) was varied from 0.1 to 1 in steps of 0.1 to directly measure the variance of open circuit voltage with respect to optical intensity variations for our specific solar panel (AM-1815CA, Sanyo, Tokyo, Japan), plotted in Supplementary Fig. S2. The most conservative percentage loss in open circuit voltage due to corresponding reduction in optical intensity was used to interpolate the voltage gain for a 13.41% optical loss, which is the difference between our 3D-printed element and anti-reflective coated PMMA injection molded design.

## Funding

The work was carried out through the finical support of Engineering and Physical Sciences Research Council grant number N032853/1, Scholarship by Ministry of Education of Saudi Arabia, Royal Academy of Engineering Research Fellowship Scheme, and Royal Academy Engineering Frontiers of Engineering for Development.

## Acknowledgments

The authors would like to thank Dr Johannes Courtial and Prof Nikolaj Gadegaard for useful conversations.

## Author contribution

REA and MJPL carried out the manufacturing, design, power simulation, data analysis and testing of the optical components. RB performed the optical modelling. The concept was devised by MJPL, and further developed in collaboration with RB and REA. The manuscript was prepared jointly by all authors.